\documentclass{article}

\usepackage[final, nonatbib]{neurips_2022}

\usepackage[utf8]{inputenc} 
\usepackage[T1]{fontenc}    
\usepackage{hyperref}       
\usepackage{url}            
\usepackage{booktabs}       
\usepackage{amsfonts}       
\usepackage{nicefrac}       
\usepackage{microtype}      
\usepackage{xcolor}         
\usepackage{graphicx} 

\title{Digital Human Interactive Recommendation Decision-Making Based on Reinforcement Learning}

\author{%
Xiong Junwu\textsuperscript{\rm 1,\thanks{Correspondence to junwu.xjw@antgroup.com}}, Xiaoyun Feng\textsuperscript{\rm 1}, YunZhou Shi\textsuperscript{\rm 2} \\
\textbf{James Zhang\textsuperscript{\rm 1}, Zhongzhou Zhao\textsuperscript{\rm 2}, Wei Zhou\textsuperscript{\rm 2}}\\
\textsuperscript{\rm 1} AntGroup\\
\{junwu.xjw, fengxiaoyun.fxy, james.z\}@antgroup.com \\
\textsuperscript{\rm 2}Damo Academy, Alibaba Group\\
\{yunzhou.syz, zhongzhou.zhaozz, fayi.zw\}@alibaba-inc.com\\
}

\begin{document}

\maketitle

\begin{abstract}
  \textit{Digital human} recommendation system has been developed to help customers find their favorite products and is playing an active role in various recommendation contexts. How to timely catch and learn the dynamics of the preferences of the customers, while meeting their exact requirements, becomes crucial in the \textit{digital human} recommendation domain. We design a novel practical \textit{digital human} interactive recommendation agent framework based on Reinforcement Learning(RL) to improve the efficiency of the interactive recommendation decision-making by leveraging both the \textit{digital human} features and the superior flexibility of RL. Our proposed framework learns through real-time interactions between the \textit{digital human} and customers dynamically through the state-of-the-art RL algorithms, combined with multi-modal embedding and graph embedding, to improve the accuracy of personalization and thus enable the \textit{digital human} agent to timely catch the attention of the customer. Experiments on real business data demonstrate that our framework can provide better personalized customer engagement and better customer experiences.
\end{abstract}

\section{Introduction}\label{sec:introduction} 
A \textit{digital human} can reinterpret customers' needs and can promptly respond to the customers' activities in the fast-growing meta-verse domain, by leveraging artificial intelligence (AI) advances. The capabilities of such a system lie not only in meeting customers' requirements but also in the appropriate human-like responses, such as facial expressions, emotions and disposition etc. These features enable \textit{digital human} to be applied in more and more apps and messengers to allow interactions with customers for better real-time recommendation consumption experiences. 

Lately, the \textit{digital human} recommendation system has been developed to help customers find their favorite products and are playing an active role in various recommendation contexts, such as purchasing, watching short videos, and making friends on social networks, but such a system also runs into the same problem of information overload (mostly from the internet) as observed in traditional recommendation systems \cite{konstan1997grouplens, breese1998empirical}. The conventional recommendation system typically works in a two-phrase pattern, i.e, first a batch of personalized item candidates are prepared and then the candidates are ranked before delivered by the recommendation engine to the customers. This pattern may well adapt to classic passive display-based recommendation contexts, with the limitation that the customer can only passively consume the prepared items and often with a single chance of action (e.g. watching or clicking only once among the recommended items). With the outbreak of COVID-19, people tend to reduce physical social interactions, which has led to increasing online shopping activities at home. To meet people's timely needs, location-based services and real-time recommendations become more critical, which drives the need for timely comprehension of the preferences of the customers, while meeting their exact requirements during their dynamic and active interactions with the \textit{digital human}. This poses an open problem that is becoming increasingly challenging and important.

Recently, research work has widely turned to formulating recommendation decision-making as a sequential decision problem to better reflect the user-system interactions. Unlike traditional recommendation approaches such as collaborative filtering\cite{konstan1997grouplens, breese1998empirical} or content-based filtering \cite{linden2003amazon}, the problem can be formulated as a Markov decision process (MDP) \cite{bellman1957markovian}, which can be solved by Reinforcement Learning (RL) \cite{kaelbling1996reinforcement, sutton1998reinforcement, szepesvari2010algorithms} that \cite{ie2019reinforcement, li2018reinforcement} has shown great advantages in handling sequential dynamic interactions while being proficient in taking into account the long-term reward. However, the applications of RL have been mainly confined to traditional passive display-based recommendation contexts \cite{afsar2021reinforcement} with limited chances of user interactions. It remains a challenging problem to timely catch the attention of a customer either with a single interaction through bandit learning \cite{slivkins2019introduction, bouneffouf2020survey, elena2021survey} in a single session or with multiple interactions through batch RL \cite{levine2020offline, prudencio2022survey}, from historical interactions between the customer and a prepared set of item candidates.

In this paper, we focus on making better use of both the \textit{digital human} features on leveraging the superior adaptability of RL to improve the efficiency of decision-making in interactive recommendations.

We propose a novel practical \textit{digital human} interactive recommendation agent framework named \textbf{MAgent} (`\textbf{M}' here stands for digital hu\textbf{M}an), enabling the \textit{digital human} agent to actively and timely catch a customer's attention during its multi-round interactions in a single session using RL algorithms. This new paradigm of recommendation context is obviously different from most existing display-based recommendation systems that are passive. Furthermore, our proposed framework can provide better-personalized customer engagement, e.g., coherent multi-modal information interactions and better customer experiences, etc. These new sights can be leveraged to better solve the above-mentioned challenging problems of active recommendations.

\section{Related Work}
\textbf{Digital Human} A \textit{digital human} is a virtual intelligent agent (or avatar) that can speak, comprehend, infer and convey emotions with a three-dimensional digital body and can perform human-like tasks through natural language dialogues with humans, which is normally composed of behavior generation and character animation building blocks for conversational simulations and training. A \textit{digital human} can make a cohesive and interactive animated performance while conversing with humans if given a set of communicative capabilities (accent, illustration, turn-taking etc.), and behavioral requests (conversation, gesticulation, stare, etc). 

With more giant internet companies and research institutes providing many kinds of platforms (such as Google VTuber or virtual YouTuber, Microsoft XiaoIce, Meta/Facebook Codec Avatar, NVidia Omniverse Avatar Toy Jensen, and Alibaba AliME\cite{li2021alime}), it is now possible to recreate interactions between natural human and much more advanced real-time virtual humans \cite{badler1997real} of large scales at very low cost. Specifically, these platforms provide a rich set of tools for the creation and operation of virtual hosts or virtual brand spokespersons for customers. These \textit{digital human}s provide customers access to a tangible human connection to the digital world in a real-time and interactive manner with friendly experiences, equipped with human-like appearances and emotions. Besides, internet services and products provided by \textit{digital human}s are available 24/7, which can further improve operational efficiency and reduce labor costs. 

\textbf{Digital Human Recommendation System} 
Internet-based \textit{digital human} characters with AI capability driven by data are being widely applied in personalized recommendation systems backed by abundant cutting-edge technologies (such as multi-modal fusion, knowledge graph and three-dimensional \textit{digital human} modeling) in content delivery. Moreover, user preferences are expressed in various modalities, such as texts, images, videos and audio data, and can be learned through multi-modal and graph data with feature embeddings. Text-based multi-modal Review Generation \cite{quoc1864multimodal} is a neural network approach that jointly models a rating prediction component and a review text generation component. An attention model was proposed in the image-based Attentive Collaborative Filtering (ACF) model \cite{chen2017attentive} that is composed of the component-level attention module that learns to select informative components of multimedia items, and the item-level attention module that learns the items' preference scores. Graph-based Knowledge Graph (KG) Attention Network (KGAT) model is proposed in \cite{wang2019kgat}, which explicitly learns the high-order connectivities in KG in an end-to-end fashion. These works seem to have been incorporated into the \textit{digital human} systems to further improve the performance of recommendations. One typical example is AliMe Avatar \cite{li2021alime} (similar to Vtuber), where a \textit{digital human} has been designed for E-commerce live-streaming retail sales recommendation service, whose core ability is to provide customers with better understanding of the products, thereby promoting the online purchases in the virtual live broadcast room.

As mentioned in Sec.~\ref{sec:introduction}, these algorithms typically consider the case where each customer can only passively consume the prepared items with a single chance of action and may well be applied to classic recommendation systems under traditional contexts in a passive manner. However, these algorithms cannot be easily adapted to fully dynamic contexts appropriately, especially in interactive recommendation decision-making contexts such as recommendations in a virtual live broadcast room.

\textbf{Reinforcement Learning based Recommendation System} RL is a trending area of machine learning focusing on how an intelligent agent or multi-agent (we consider only one agent in our work here) takes actions in a dynamic environment to maximize the cumulative reward. Specifically, the agent learns what to do (how to map situations to actions) to maximize the reward \cite{sutton1998reinforcement}, which is complicated by the problem of learning user behaviors through trial-and-error interactions in a dynamic environment \cite{kaelbling1996reinforcement}. To improve the efficiency and performance of traditional recommendation systems, various behavior-based methods have been proposed. The collaborative filtering method SVDFeature \cite{chen2012svdfeature} is designed to solve the collaborative feature-based matrix factorization effectively. Deep learning-based recommendation algorithms Convolutional Matrix Factorization (ConvMF) \cite{kim2016convolutional} is a novel context-aware recommendation model ConvMF that integrates a convolutional neural network (CNN) into probabilistic matrix factorization (PMF), which captures contextual information of documents and further enhances the prediction accuracy of the ratings. \cite{ma2008sorec} proposes a factor analysis approach based on probabilistic matrix factorization to solve the problems of data sparsity and poor prediction accuracy by employing both user's social network information and rating records. Unlike these traditional recommendation algorithms, RL is able to handle the sequential dynamic user-system interaction with better long-term user engagement \cite{afsar2021reinforcement}. However, as mentioned in Sec.~\ref{sec:introduction}, the applications of RL have been mainly confined to traditional passive display-based recommendation contexts, i.e., although they have the same motivation as our method, the efficiency problem in such context is not explicitly addressed.

\section{Background} 
With customized appearances, multi-modal interactions and the reactive product broadcasting, \textit{digital human}'s  on-the-fly recommendations based on RL can improve the performance of the system through learning from dynamics between the \textit{digital human} agent and customers. In classical passive display recommendation systems, many more items, limited by display resources, have little chance to be displayed (i.e., to be visible to customers) and customers usually have few interactions, such as only a few clicks or even no clicks, if they have no interests.

Before proceeding with the introduction of formulations and models, we introduce the following assumptions for the communication modes between the \textit{digital human} and their customers being served, i.e., one-to-one (i.e. one \textit{digital human} communicates with and services one customer in a session), one-to-many (i.e. one \textit{digital human} communicates with and services many customers simultaneously), and many-to-many (many \textit{digital human}s communicates with and services many customers), etc. In this work, we mainly focus on the typical context of the one-to-many communication mode, which has been extensively applied in many practical contexts and it is not difficult to verify that our work can be easily extended to other service modes.

In the general one-to-many communication mode of live-streaming broadcast recommendation solutions, the \textit{digital human} can naturally be formulated as an RL agent and can even play the role of a real human seller, meeting various needs during real-time interactions with customers. In the virtual live broadcast room under this mode, if we formulate an algorithm model as in the classical passive display-based recommendation system, the \textit{digital human} usually can only deliver the same products or services to its serviced customers in a session without focusing on dynamic interactions. Obviously, different users who enter the live broadcast room in the same period may have distinct demands and each customer can describe his/her requirements more clearly through multi-round interactions with the \textit{digital human}, e.g., some customers want to know the quality of the products, some focus on the features of the products, others want to learn whether there is any discount for the product and so on.

To improve the efficiency of decision-making of the \textit{digital human} interactive recommendation system, similar to meeting the personalized needs in the live broadcast room at the same time, we propose a novel Reinforcement Learning (RL) algorithm-based framework, which learns through real-time interactions between the \textit{digital human} and customers' dynamic embeddings with multi-modal and graph embeddings to improve the accuracy of the personalization and thus the efficiency of decision-making of the recommendation, which means in the long run, customers are provided with better products and improved services.

\section{Digital Human Interactive Recommendation Decision-Making Based on Reinforcement Learning}\label{sec:theoretical_algs_and_sys_frame}
In this section, we detail the model and framework of a \textit{digital human} interaction recommendation system. 

\begin{figure}
\centering
\includegraphics[width=1\textwidth]{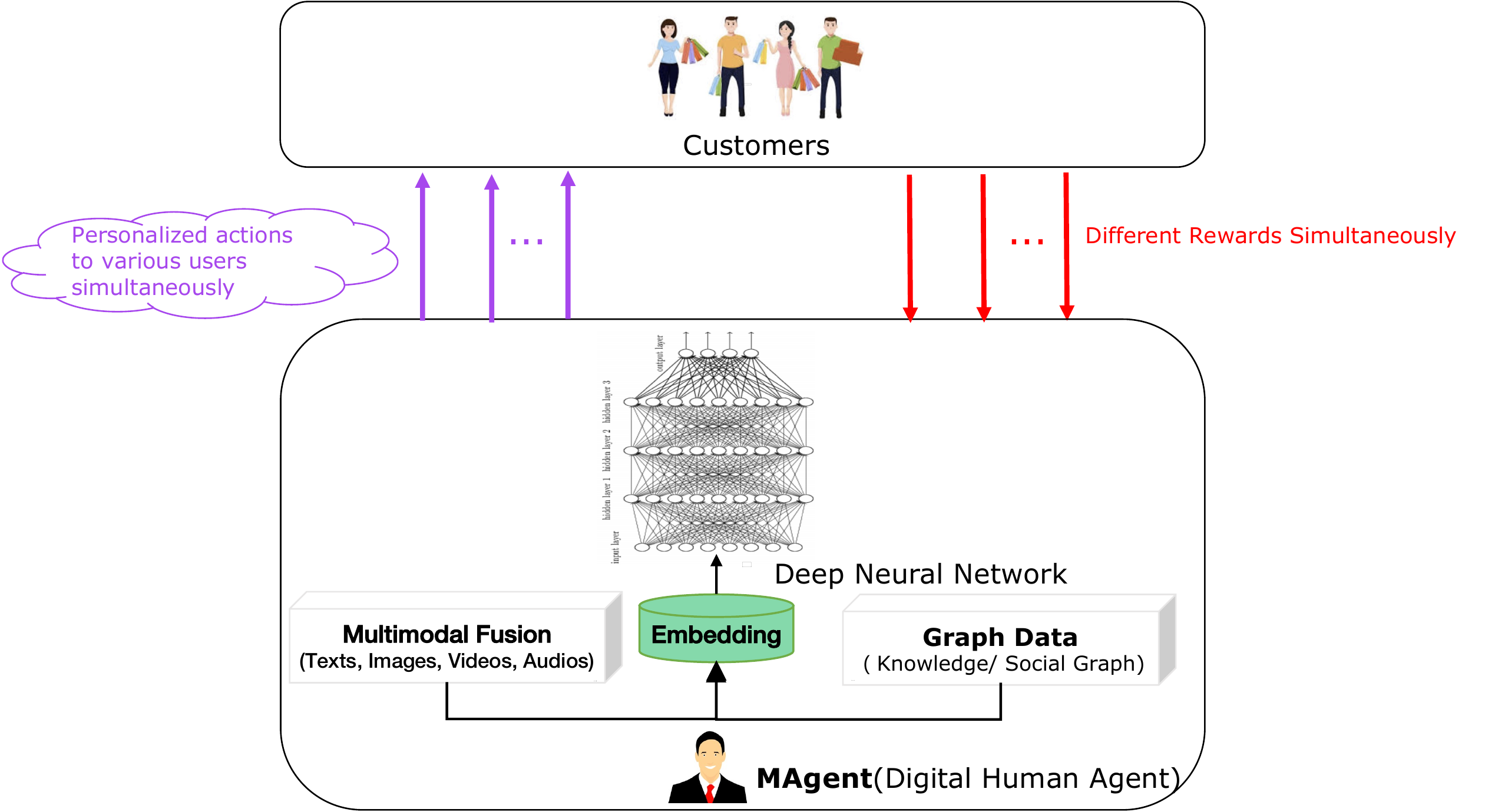}
\caption{Digital human interactive recommendation based on reinforcement learning framework.}
\label{fig:MAgent}
\end{figure}

It is straightforward and sensible to formulate the digital agent as an RL agent, due to the sequential and dynamic interactions, as well as the long-term goal of user engagement. The goal for general RL algorithms is to learn a behavior policy directly, i.e., to maximize the long-term cumulative reward expectation value: 
\begin{equation}
\pi^* =\arg\max_{\pi}\mathbb{E}_{(s_t,a_t)\sim\rho}[\sum_t R(s_t,a_t)]
\end{equation}
Here, we introduce the maximum entropy-based RL algorithm \cite{haarnoja2018soft} for its stable performance in various contexts. The maximum entropy-based RL, in addition to the above goals, also maximizes the entropy of each output of the policy: 
\begin{equation}
\pi^*=\arg\max_{\pi}\mathbb{E}_{(s_t,a_t)~}[\sum_t \underbrace{R(s_t,a_t)}_{reward}+\alpha\underbrace{H(\pi(\cdot|s_t))}_{entropy}]
\end{equation}
Our proposed Reinforcement Learning (RL) algorithm-based framework for \textit{digital human} interactive recommendation can therefore be described as in Figure \ref{fig:MAgent}.

The advantages of our novel framework can be summarized as follows. Through the introduction of RL, the \textit{digital human} becomes the agent of the RL framework, and thus can naturally act as the role of an actor to meet varying needs during real-time interactions with different customers. Moreover, this framework can leverage various state-of-the-art RL algorithms to further improve the accuracy of personalization by adapting to different statistical natures or contexts of the real-world applications, and thus can provide the customers with better products and services. For example, for small-scale customer sizes, on-policy RL algorithms such as SARSA (Equation \ref{eq_td_sarsa}) \cite{sutton1998reinforcement} can be adopted and SlateQ (Equation \ref{eq_slateq})\cite{ie2019reinforcement} algorithm can be introduced for large-scale systems, where a slate can be used as mini-batch actions to different groups of customers.
TD(0)/SARSA Q function: 
\begin{equation}\label{eq_td_sarsa}
Q^{\pi} = R(s,A) +\gamma\sum_{s\in S}P(s'|s,A)V_{\pi}(s') 
\end{equation}
SlateQ with the full decomposition of the on-policy Q-value function: 
\begin{equation}\label{eq_slateq}
Q^{\pi} = R(s,A) +\gamma\sum_{s\in S}P(s'|s,A)V_{\pi}(s')\\
= \sum_{i\in A} P(i|s,A) \overline{Q}^{\pi}(s,i)
\end{equation}
Here, $\overline{Q}^{\pi}(s,i) = R(s,i) +\gamma\sum_{s\in S}P(s'|s,i)V_{\pi}(s')$ is the Expected LTV of clicked (or converted) item $i$ and $P(i|s,A)$ is User of a user choice model (pCTR, pCVR ect.).

Besides the advantages mentioned above, RL framework can be embedded with multi-modal and graph embeddings. Multi-modal embedding (similar to live-streaming broadcast context) can fuse multi-modal data during the interactions between the \textit{digital human} and customers, while graph embedding can build knowledge graphs \cite{xu2021alime}, providing a cognitive profile for products, for a better understanding by customers.

\section{Experiments}
In this section, we carry out the following experiments or A/B testing to validate the performance of our proposed algorithm framework \textbf{MAgent} and compare it against the state-of-the-art algorithm running as a typical passive display-based recommendation system. Specifically, we validate the performances of RL algorithm \textbf{SAC} (Soft actor-critic) \cite{haarnoja2018soft} against \textbf{DFM} (Deep factorization model) \cite{guo2017deepfm} to evaluate the performance of \textit{digital human}'s recommendation decision-making under the context of live-streaming broadcast with real-world business data and compare the corresponding conversion rate of transactions on regular days, as well as on marketing campaign days through metrics of MRR (Mean reciprocal rank)\cite{le2010optimization} and Hits@K, etc.

To meet the varying demands of the customers timely in the live broadcast room, we leverage the dynamic learning capability of RL. Through personalized learning \textit{digital human} agents, customers who enter the virtual live broadcast room can be better served, and the digital service provider platform can reduce the loss of customers and thus improve the conversion rate of transactions. Please note that the transaction conversion rate on marketing campaign days is usually much higher than that on regular days. According to data statistics in a real business scenario, the average number of times that customers enter the live broadcast is very few, while the interactions between customers and the \textit{digital human} agent are dynamic and sequential in time. Our sample datasets are built from customers' interaction and transaction behaviors in a virtual live broadcast. For each customer who enters the virtual live broadcast room, different content is displayed with multi-round interactions. The sample dataset is divided into a training set and a validation set w.r.t. the time dimension. The features of \textbf{SAC} are defined as follows:
\begin{itemize}
\item Action: Action field includes exposure content types (8 types in total).
\item Context: Context field contains two parts. The first part is user-related (e.g., user ID/identification who enters the channel of the live broadcast room etc.); and the second part is merchant-related (e.g., the store ID that the user interacts with).
\item State: The number of exposures of different content types of customers, including contents of discount cards, selected pictures on detail pages, shopping content, product cards, and historical live broadcast slices, etc.
\item Next State: The number of exposures of the content type corresponding to the action plus 1.
\item Reward: If the customer has a behavior of clicks or making a deal, the corresponding reward is 1, otherwise the reward is 0. 
\end{itemize}

\begin{table}
\caption{Digital human recommendation performance based on RL framework}
\label{table:perf}
\centering
\begin{tabular}{|c|c|c|}
\hline
& MRR & Hit1@Recall \\\hline
\textbf{DFM} & 0.255 & 3.9\%\\\hline
\textbf{SAC} & 0.308 & 4.2\%\\\hline
\end{tabular}
\end{table}

Our experiment results are shown in Table \ref{table:perf}, where the baseline model is a transaction prediction model based on \textbf{DFM}, and experimental data demonstrate that compared with a classical passive display-based recommendation system, our proposed algorithm framework significantly improves the performance in the task of product's content-type decision-making through dynamic learning from customers' interactive behaviors with the \textit{digital human} agent. Here, we only compare the recommendation decision-making on the product content ID level for RL and the online display is based on the granularity of content ID too. The overall decision elements are shown as follows:
\begin{itemize}
\item Content ID: Total number of content IDs is around ten thousand. As the store expands, contents might be updated, and the set of content IDs changes dynamically.
\item Content tab: The content tab includes commodity display, color test, and detailed display, etc. Only some of the content tabs have current labels.
\item Content type: There are 7 content types, including commodity cards, video slices, and shopping, etc.
\end{itemize}

Obviously, there is still huge potential improvement to be explored and we will have a much more comprehensive comparison in a larger combinatorial decision space in our future work, and it will be beneficial for the \textit{digital human} service provider to improve user activities and promote the conversion rate of transactions in the virtual live broadcast room in various practical business domains.

\section{Discussion}\label{sec:discussion}
Different from traditional recommendation algorithms, we proposed a novel Reinforcement Learning (RL) framework to improve the accuracy of personalization through real-time interactions between the \textit{digital human} and customers dynamically and thus better satisfied the real-time requirements of customers. Besides, our RL framework can be embedded with multi-modal and graph embedding to further improve the accuracy of key metrics in \textit{digital human} recommendation. Various experiments have been carried out and proved that it could meet the long-term value of platform development. However, we have to point out that if customers are not interested in or even feel unpleasant about the appearance of the \textit{digital human} when they are shopping online, relevant recommendations would bring uncomfortable effects. We will carry out relevant research to mitigate or even avoid corresponding negative effects in our future work.

Other \textit{digital human} services such as personalized contexts advertisement and search systems can also benefit from similar framework designs.

In the future, we will build this framework with the latest real-time NoSQL database such as HBase \cite{hbase2009apache} similar to real-time recommendation framework in \cite{xiong2019recommendation} and real-time or high-efficiency AI system, such as Ray\cite{moritz2018ray} or Flink\cite{carbone2015apache}, to further improve the efficiency of the end-to-end sampling, training and online inferences.
\bibliographystyle{unsrt} 
\bibliography{ref} 

\newpage
\section*{Checklist}
\begin{enumerate}
\item For all authors...
\begin{enumerate}
\item Do the main claims made in the abstract and introduction accurately reflect the paper's contributions and scope?
\answerYes{}
\item Did you describe the limitations of your work?
\answerYes{See Section~\ref{sec:discussion}}
\item Did you discuss any potential negative societal impacts of your work?
\answerYes{See Section~\ref{sec:discussion}}
\item Have you read the ethics review guidelines and ensured that your paper conforms to them?
\answerYes{}
\end{enumerate}
\item If you are including theoretical results...
\begin{enumerate}
\item Did you state the full set of assumptions of all theoretical results?
\answerYes{See Section~\ref{sec:theoretical_algs_and_sys_frame}}
\item Did you include complete proofs of all theoretical results?
\answerYes{See Section~\ref{sec:theoretical_algs_and_sys_frame}}
\end{enumerate}
\item If you ran experiments...
\begin{enumerate}
\item Did you include the code, data, and instructions needed to reproduce the main experimental results (either in the supplemental material or as a URL)?
\answerNo{The code and the data are proprietary.}
\item Did you specify all the training details (e.g., data splits, hyperparameters, how they were chosen)?
\answerNo{The code and the data are proprietary.}
\item Did you report error bars (e.g., with respect to the random seed after running experiments multiple times)?
\answerYes{In table \ref{table:perf}}
\item Did you include the total amount of compute and the type of resources used (e.g., type of GPUs, internal cluster, or cloud provider)?
\answerNo{The code and the data are proprietary.}
\end{enumerate}
\item If you are using existing assets (e.g., code, data, models) or curating/releasing new assets...
\begin{enumerate}
\item If your work uses existing assets, did you cite the creators?
\answerNo{}
\item Did you mention the license of the assets?
\answerYes{The license is Apache 2.0}
\item Did you include any new assets either in the supplemental material or as a URL?
\answerNA{The data are proprietary.}
\item Did you discuss whether and how consent was obtained from people whose data you're using/curating?
\answerNA{The data are proprietary.}
\item Did you discuss whether the data you are using/curating contains personally identifiable information or offensive content?
\answerNA{The data are proprietary.}
\end{enumerate}
\item If you used crowdsourcing or conducted research with human subjects...
\begin{enumerate}
\item Did you include the full text of instructions given to participants and screenshots, if applicable?
\answerNA{}
\item Did you describe any potential participant risks, with links to Institutional Review Board (IRB) approvals, if applicable?
\answerNA{}
\item Did you include the estimated hourly wage paid to participants and the total amount spent on participant compensation?
\answerNA{}
\end{enumerate}
\end{enumerate}
\appendix
\end{document}